\documentclass[prb, reprint]{revtex4-1}
\usepackage{graphicx}
\usepackage{amsmath}
\usepackage{subcaption}

\begin{document}

\title[]{Efficient electron open boundaries for simulating electrochemical cells}

\author{Mario G. Zauchner}
\homepage{mario.zauchner15@imperial.ac.uk}
\affiliation{Department of Materials and Thomas Young Centre,
Imperial College London, South Kensington Campus, London SW7 2AZ, U.K.}%

\author{Andrew P. Horsfield}
\homepage{a.horsfield@imperial.ac.uk}
\affiliation{Department of Materials and Thomas Young Centre,
Imperial College London, South Kensington Campus, London SW7 2AZ, U.K.}%

\author{Tchavdar N. Todorov}
\homepage{t.todorov@qub.ac.uk}
\affiliation{Atomistic Simulation Centre, School of Mathematics and Physics, Queen's University Belfast, Belfast BT7 1NN, U.K.}%

\date{\today}

\begin{abstract}
Non-equilibrium electrochemistry raises new challenges for atomistic simulation: we need to perform molecular dynamics for the nuclear degrees of freedom with an explicit description of the electrons, which in turn must be free to enter and leave the computational cell. Here we present a limiting form for electron open boundaries that applies when the magnitude of the electric current is determined by the drift and diffusion of ions in solution, and which is sufficiently computationally efficient to be used with molecular dynamics. We demonstrate its capabilities by way of tight binding simulations of a parallel plate capacitor with and without a dimer situated in the space between the plates.
\end{abstract}

\maketitle

\section{\label{sec:level1}Introduction\protect}

A traditional electrochemical experiment involves at least two electrodes immersed in an aqueous solution \cite{Bockris1973}, with the reactions that take place depending on the bias applied between those electrodes. This arrangement can apply to batteries, electrolysis, and even corrosion. Two key features of this system are: electrons can flow onto and off the electrodes; the reactions occur when the system is out of equilibrium. To simulate this system we need to capture these two features.

An appropriate simulation methodology thus must be able to combine open boundary conditions for the electrons with molecular dynamics (MD) for the solution. This puts very tight efficiency constraints on the open boundary formalism as even MD performed with ground state electronic structure methods can be computationally very demanding. Here we present an open boundary formalism appropriate for electrochemical problems that has essentially the same computational cost as ground state electronic structure methods.

We note that important insights have been arrived at by means of equilibrium simulations which can address aspects of the problem. For example, the variation of interface properties with electron chemical potential can be investigated by adding electrons to the computational cell \cite{rossmeisl2008}, and dynamic simulations to determine the symmetry factor associated with the detachment of an ion have been performed \cite{Drechsel-Grau2015}.

While there are several well tested codes for open boundaries \cite{Brandbyge2002,Rocha2004,Ozaki2010,ATK}, these tend to be computationally too expensive for non-equilibrium MD. The Hairy Probes (HP) method \cite{HairyProbes2016,McEniry2007} has previously been shown to be an efficient solution for electron open boundary simulations of nano-scale systems. However, even the standard HP method is not efficient enough for MD over larger time scales. Fortunately, for the specific case of electrochemistry we can exploit the fact that the current is carried by ions in solution, rather than ballistic electrons, to produce a simplified version of HP. This method remains mathematically well defined, while having much higher efficiency, and is straightforward to incorporate into existing electronic structure codes.

As current transport occurs through the drift and diffusion of ions in solution, we can consider the limit where electrons enter and leave the electrodes slowly. This corresponds to the use of the HP formalism for weakly coupled probes. It was speculated in an earlier paper that this might be a possible approach \cite{HairyProbes2016}; here we illustrate how the method works in practice.

In this limit we regain the familiar single particle picture with molecular orbitals populated by electrons \cite{HairyProbes2016}, but the population of a given level is now controlled by the attached probes and their associated electrochemical potentials. This allows different electrochemical potentials to operate within the system, producing non-equilibrium conditions, while retaining the efficiency of traditional ground state electronic calculations.

We have implemented this method in the Tight Binding (TB) package PLATO \cite{Horsfield1997,Kenny2009} (Package for Linear-combination of Atomic Type Orbitals). We have investigated a parallel plate capacitor consisting of two Cu plates, viewing this as the simplest approximation to an electrochemical cell we could imagine. We use a simple Empirical Tight Binding (ETB) model to study this system, and so inspect the correctness (or otherwise) of the method. We performed static relaxation calculations of the atomic coordinates of the system under an applied voltage, to show that the forces generated by the method using the expression from Ehrenfest Dynamics \cite{Horsfield2004,HairyProbes2016} are sufficiently close to the derivative of the expression used for the total energy to allow energy conserving MD simulations. We were then able to compute the bond length and charge distribution on a copper dimer between the two capacitor plates as a function of applied bias; atoms in the capacitor plate as well as the dimer were free to move. Finally, we performed MD simulations in the absence of thermostats for the dimer between charged plates to illustrate the applicability to MD. For closed boundary simulations these would be constant energy. We find that in this case energy is remarkably well conserved, even with the presence of open boundaries.

\section{Formalism}

The method used here is based on the HP formalism, described in detail in Horsfield \emph{et al} \cite{HairyProbes2016}. The central idea is that probes are attached to individual atomic orbitals on atoms, and that these probes are coupled to electron reservoirs characterized by an electron temperature and electrochemical potential. Following the scattering theory arguments of Todorov \cite{Todorov1993} the following expression for the single particle density matrix can be derived
\begin{equation}
\rho_{\beta\beta'}=2\sum_{rs}f_{rs}\chi_{\beta}^{(r)}\chi_{\beta'}^{(s)*}\label{eq:rho}
\end{equation}
where the factor of 2 is for spin degeneracy and
\begin{equation}
f_{rs}=\frac{1}{2\pi} \sum_{p} \Gamma_p \zeta^{(r)*}_{\beta_p} \zeta^{(s)}_{\beta_p} \int_{E_{p, c}}^{\infty} \frac{f^{(p)}(E)}{(E-\epsilon^{(r)})(E - \epsilon^{(s)*})} \mathrm{d}E
\label{limocc}
\end{equation}
can be thought of as a generalised occupancy. Here $\zeta^{(r)}_{\beta}$ and $\chi_{\beta}^{(r)}$ are left and right eigenstates respectively given by
\begin{eqnarray}
\sum_{\beta'}\left[H_{\beta\beta'}-\delta_{\beta\beta'}\sum_{p}\frac{\mathrm{i}}{2}\Gamma_{p}\delta_{\beta\beta_{p}}\right]\chi_{\beta'}^{(r)} & = & \epsilon^{(r)}\sum_{\beta'}S_{\beta\beta'}\chi_{\beta'}^{(r)}\label{eq:eigen}\\
\delta_{rs} & = & \sum_{\beta\beta'}\zeta_{\beta}^{(r)*}S_{\beta\beta'}\chi_{\beta'}^{(s)}\label{eq:orthonormal}
\end{eqnarray}
where $\epsilon^{(r)}$ is the corresponding eigenvalue. The subscript $\beta$ indexes atomic orbitals, $\Gamma_p$ is the coupling strength of probe $p$, $f^{(p)}$ is the Fermi function with temperature end electron chemical potential for that probe, and $\beta_p$ is the orbital to which the probe is attached. The quantities $H_{\beta\beta'}$ and $S_{\beta\beta'}$ are the Hamiltonian and overlap matrices respectively.

In the limit of small coupling $f_{rs}$ becomes diagonal with:
\begin{equation}
f_{rr}\rightarrow \frac{1}{2\pi} \sum_{p} \Gamma_p \lvert\zeta^{(r)}_{\beta_p}\rvert^2  \int_{E_{p, c}}^{\infty} \frac{f^{(p)}(E)}{\lvert E-\epsilon^{(r)}\rvert^2} \mathrm{d}E
\end{equation}
If we rewrite $\epsilon^{(r)}$ as $\varepsilon^{(r)} - i \eta^{(r)}$ we get:
\begin{align}
f_{rr}\rightarrow\frac{1}{2\pi}\sum_{p}\Gamma_p \frac{\lvert\zeta^{(r)}_{\beta_p}\rvert^2}{\eta^{(r)}}\int_{E_{p, c}}^{\infty} \frac{f^{(p)}(E) \eta^{(r)} }{\lvert E-\varepsilon^{(r)}\rvert^2 + \eta^{(r)2}} \mathrm{d}E \\
\rightarrow\sum_{p}\frac{1}{2\eta^{(r)}} \Gamma_p \lvert\zeta^{(r)}_{\beta_p}\rvert^2  f^{(p)}(\varepsilon^{(r)})
\label{frr}
\end{align}
From first order perturbation theory we have $\eta^{(r)} = \frac{1}{2}\sum_{p}\Gamma_p\lvert\zeta^{(r)}_{\beta_p}\rvert^2$. Substituting this into Eq. \ref{frr}, and assuming $\Gamma_p$ is independent of $p$, finally yields \cite{HairyProbes2016}:
\begin{equation}
f_r = f_{rr} \rightarrow \frac{\sum_{p} \lvert\zeta^{(r)}_{\beta_p}\rvert^2 f^{(p)}(\varepsilon^{(r)}) }{\sum_{p} \lvert\zeta^{(r)}_{\beta_p}\rvert^2 }
\label{fr}
\end{equation}
which can be interpreted as the weighted average of Fermi functions from each probe.

Eq. \ref{fr} works very well for strongly bonded systems, such as metallic plates. However, in electrochemistry we encounter systems where solvent molecules are only loosely connected to the electrodes, which makes it difficult for electrons to reach solution molecules. A simple solution to this problem can be achieved by making $\Gamma_p$ dependent on $p$ again, and separating the probes into the main probes $p$ with coupling strength $\Gamma$ that impose the voltage, and solution probes $s$ with coupling strength $\alpha\Gamma$, where $\alpha\ll1$. If the Fermi function for the solution probes is $\bar{f}(\epsilon)$, then the expression for the occupancy becomes:
\begin{equation}
f_{r} =  \frac{\alpha \bar{f}(\varepsilon^{(r)})\sum_{s} \lvert\zeta^{(r)}_{\beta_s}\rvert^2 + \sum_{p} \lvert\zeta^{(r)}_{\beta_p}\rvert^2 f^{(p)}(\varepsilon^{(r)}) }{\alpha\sum_{s} \lvert\zeta^{(r)}_{\beta_s}\rvert^2 + \sum_{p} \lvert\zeta^{(r)}_{\beta_p}\rvert^2 }
\label{eq:fr_alpha}
\end{equation}

If the main probes do not couple at all to the state $r$ then we get $f_r=\bar{f}(\epsilon^{(r)})$, which is the result for a ground state calculation, and is independent of $\alpha$. We require $\alpha$ to be small so that when the main probes are coupled, their contribution to the level occupancy dominates that from the solution reservoir, ensuring a bias can be applied as desired.

The energy of a system is not well defined when we have open boundaries. However, in this simplified formalism we compute energy using the same expression as for ground state calculations, but with the molecular orbital populations given by Eq. \ref{eq:fr_alpha}. The forces are evaluated using the expression
\begin{equation}
F_\nu = -\mathrm{Tr}\left\{\rho\frac{\partial H}{\partial R_\nu}\right\} - \frac{\partial\Phi}{\partial R_\nu}
\label{eq:force}
\end{equation}
where $F_\nu$ is a component of the atomic forces ($\nu$ combines the atomic index and direction), $R_\nu$ is a component of the atomic positions, $\rho$ is the single particle density matrix, $H$ is the single particle Hamiltonian, and $\Phi$ combines the nuclear-nuclear interaction and double counting terms. This expression corresponds to treating the forces as originating from an Ehrenfest Dynamics calculation \cite{Horsfield2004}.

\section{Method}

The limiting case of the HP method for weakly coupled probes was added to the TB software PLATO. To study the behaviour of the new algorithm, we investigated a parallel plate capacitor consisting of a total of 256 Cu atoms (128 per plate), with 64 probes attached to the outer atoms of each plate. The solution probes with coupling of $\alpha\Gamma$ are then applied to all other atoms (in this case, the two layers in the plates facing the opposite plates). We used an orthogonal tight binding model by Sutton {\it et al}. \cite{Sutton2001}, where one s orbital is assigned to each atom. Each atom is allowed to acquire a monopole charge, with the Coulombic interaction being allowed to extend to infinity; the on-site repulsion is given by $U = 6.80$ eV. All probes have vanishingly small coupling strength and a temperature of $k_b T_p=$ 0.0136 eV. The temperature enters through the Fermi function assigned to each probe, and determines the electron population associated with the probe.

In the first step of the simulation, a single arbitrary electrochemical potential is assigned to every probe. The algorithm then adjusts this one chemical potential until the system as a whole is charge neutral. The determined chemical potential is then used as the reference chemical potential of the system. 

\begin{figure}[h]
\captionsetup[subfigure]{position=top}
\begin{subfigure}{\columnwidth}
\includegraphics[width=1\textwidth]{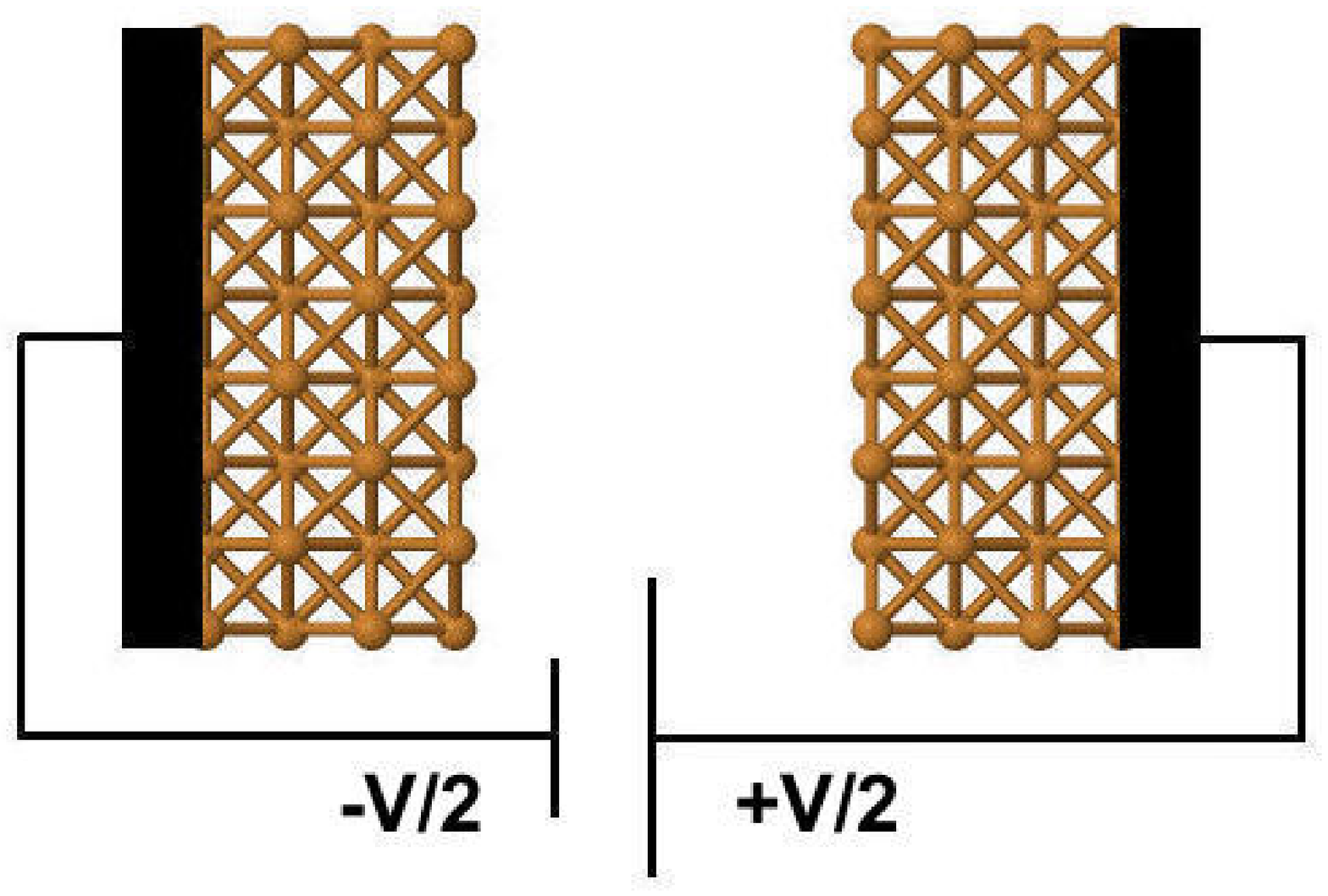}
\vspace{-1\baselineskip}
\caption{}
\vspace{-0.7\baselineskip}
\label{setup:sub1}
\end{subfigure}
\begin{subfigure}{\columnwidth}
\includegraphics[width=1\textwidth]{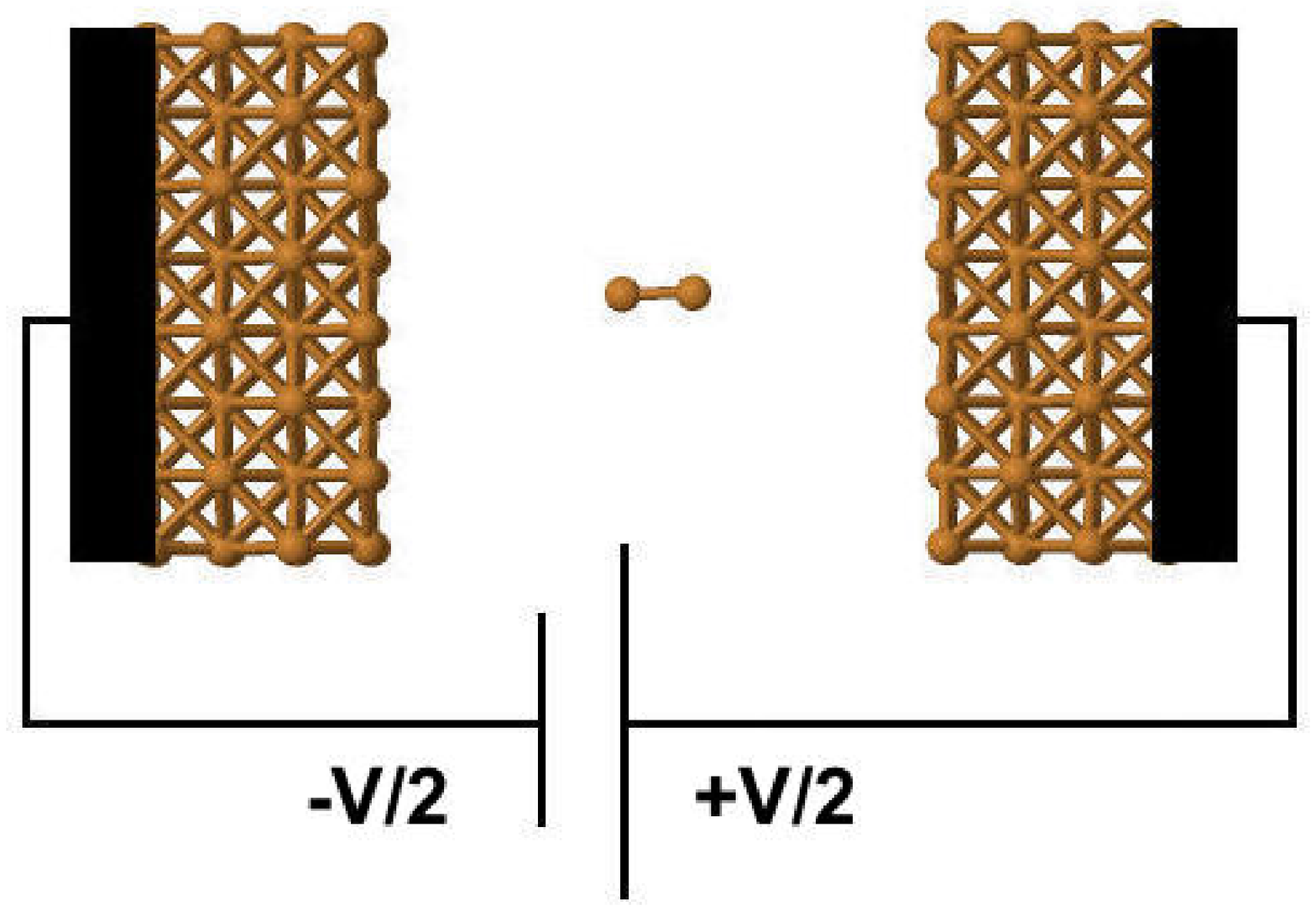}
\vspace{-1\baselineskip}
\caption{}
\vspace{-1\baselineskip}
\label{setup:sub2}
\end{subfigure}
\caption{
Capacitor setup for the Hairy Probes calculations. There is a total of 256 Cu atoms (128 per plate), with 64 probes attached to the outer two planes of atoms on each plate. a) The setup for determination of the charge distribution, and the net charge as a function of applied voltage. b) The setup for the bond length calculations. Note that the separation of the plates was doubled relative to the previous case to ensure the dimer is not connected to the plates.}
\label{setup}
\end{figure}

In the next step of the simulation a bias is applied, which leads to an anti-symmetric shift in the chemical potential of the two terminals: half the bias is added as a positive shift in the electron chemical potential to the probes attached to one plate, while half is added as a negative shift to the other plate. By varying the applied bias we determined the net charge on one plate as a function of applied bias. The setup is illustrated in Fig. \ref{setup:sub1}. Note that the plates are far enough apart that there is zero electron hopping between them. To maintain global charge neutrality, we adjust the reference electrochemical potential once a bias is applied.

The process was then repeated for the system shown in Fig. \ref{setup:sub2} in which a copper dimer has been added between the plates. The bond length of the copper dimer was computed for different voltages. Note that the separation between the copper plates was doubled to ensure the dimer does not form a bond with the copper plates.

Finally, the relaxed system with an applied bias of $0.408$ V was used to perform a simple molecular dynamics simulation.

\section{Results}

\subsection{Parallel plate capacitor}

\begin{figure}[h!]
\includegraphics[width=\columnwidth]{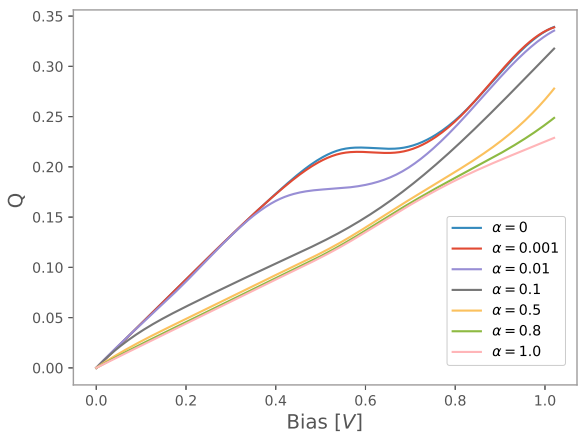}
\caption{The total charge accumulated on a capacitor plate for different voltages for a parallel plate capacitor. The capacitor consists of 256 Cu atoms, 128 atoms per plate. A range of values of the solution probe coupling $\alpha$ is considered. Note the shoulder for small values of $\alpha$.}
\label{dQ_V}
\end{figure}

\begin{figure}[h!]
\includegraphics[width=\columnwidth]{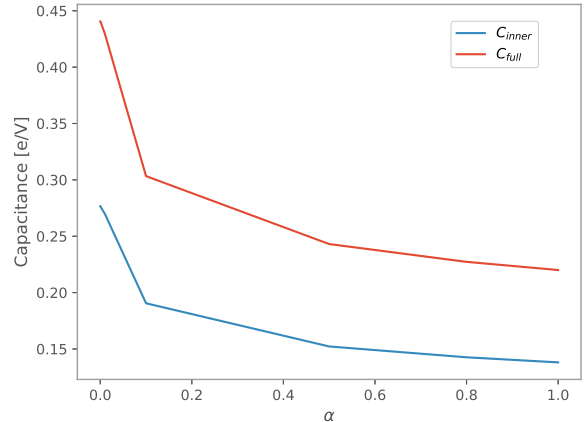}
\caption{Dependence of capacitance on solution probe coupling. For the red (upper) line, the capacitance is given by the total charge on a plate divided by the applied voltage. For the blue (lower) line the capacitance is defined as the charge on the half of the plates closest to the opposite plates, divided by the applied voltage. }
\label{capacitances}
\end{figure}

For a range of values of the coupling coefficient $\alpha$, we determined the total charge on one plate as a function of the applied bias. From the capacitor law $Q=CV$ (where $Q$ is the charge on a plate, $C$ is the capacitance, and $V$ is the applied voltage), one would expect a linear dependence of $Q$ on $V$. From the slope of a line we can determine the capacitance as a function of $\alpha$. From Fig. \ref{dQ_V} we see that there is a plateau for low coupling strengths. This occurs because the left and right chemical potentials have entered regions of low density of states. Once they re-enter regions of higher density of states the charge continues to increase with a similar slope as before. For the determination of the capacitance, we fit a straight line to the region before the plateau. The capacitance as a function of solution probe coupling is shown in Fig. \ref{capacitances}. The lowering of capacitance as $\alpha$ increases can be explained as follows. As $\alpha$ increases, the relative contribution of the main probes (which have a finite bias applied) is lowered, while that from the solution probes (with zero applied bias) increases, leading to a decrease in the effective bias applied experienced by the plates. Thus we need to apply a higher external bias to generate a given charge, leading to a reduced capacitance.

\begin{figure}[h!]
\includegraphics[width = \columnwidth]{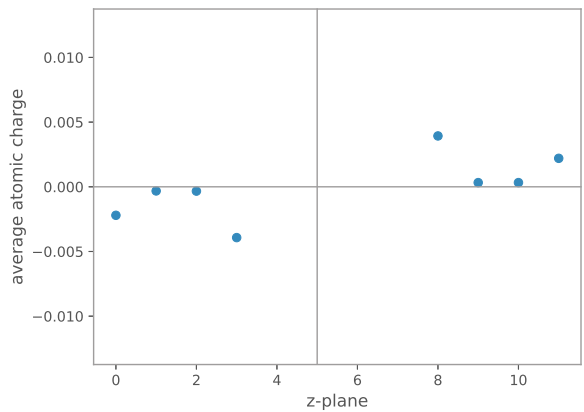}
\caption{Average charge per atom on each plane of atoms in the capacitor plates at 0.408 V bias and solution coupling strength $\alpha$ = 0.0. Note the increased charge at the outermost and innermost layers of the plates.}
\label{dQ}
\end{figure}

In Fig. \ref{dQ} we show the charge distribution inside the capacitor. We see that there is a concentration of charge on the innermost and outermost planes of atoms. This suggests two definitions of the capacitance, with the two lines in Fig. \ref{capacitances} corresponding to these two definitions. One is the usual definition of capacitance $C = Q/V$ where $Q$ is the total charge on the plate, and $V$ the applied voltage between the plates. The other definition is $C = Q_{\mathrm{inner}}/V$ where $Q_{\mathrm{inner}}$ is the charge on the two planes in a plate nearest the opposite plate.

This second definition is interesting as it leads to better agreement with the standard parallel plate capacitor expression $C = \epsilon_0 A/d = 0.16$ e/V where $\epsilon_0$ is the permittivity of free space, $A$ is the area of the capacitor plate, and $d$ is the separation between the plates. The better agreement follows from the derivation of this standard expression: it assumes charge distributed over one face of each plate.

\begin{figure}[h!]
\includegraphics[width=\columnwidth]{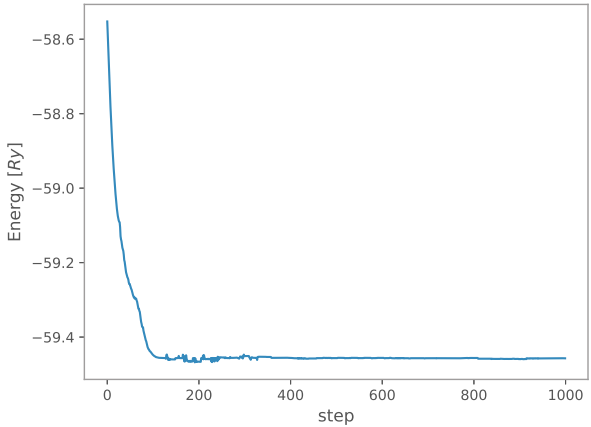}
\caption{Evolution of the energy of the capacitor upon relaxation under an applied bias of 0.816 V.}
\label{energy}
\end{figure}

Under an applied bias of 0.816 V, we relaxed the forces on the atoms in the capacitor as shown in Fig. \ref{setup:sub1}. We find that the forces are accurate enough to lower the system's energy in a smooth fashion upon relaxation, as shown in Fig. \ref{energy}. We see that the energy decreases sharply over the first 100 steps, and then remains relatively constant at the equilibrium value.

\subsection{Bond length of copper dimer under applied bias}

\begin{figure}[h!]
\captionsetup[subfigure]{position=top}
\begin{subfigure}{\columnwidth}
\includegraphics[width=1\textwidth]{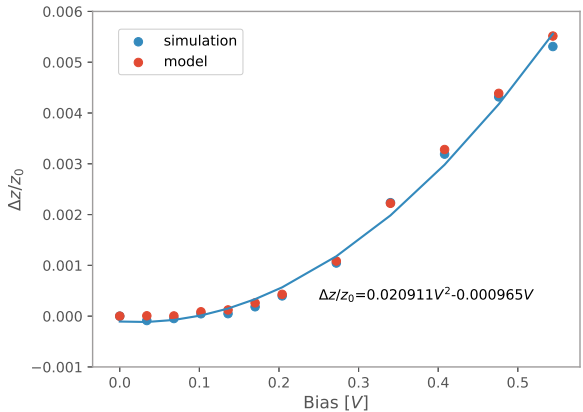}
\vspace{-2\baselineskip}
\caption{}
\label{bondlength:sub1}
\vspace{-0.7\baselineskip}
\end{subfigure}
\vspace{-0.1\baselineskip}
\begin{subfigure}{\columnwidth}
\includegraphics[width=1\textwidth]{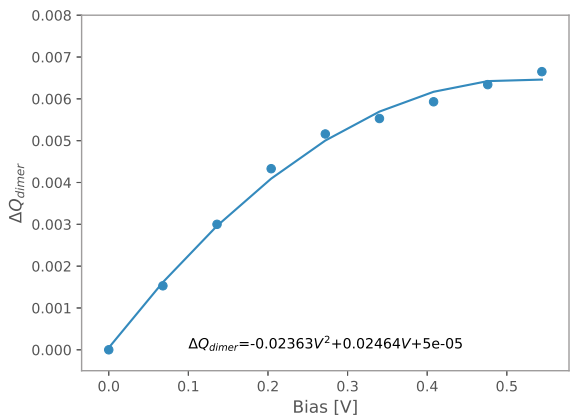}
\vspace{-2\baselineskip}
\caption{}
\label{bondlength:sub2}
\vspace{-0.2\baselineskip}
\end{subfigure}
\caption{a$)$ The fractional change in bond length as a function of bias. The blue dots are from the simulation, the red dots are from our simple model, and the line is a quadratic fit. 
b$)$ The charge difference between the dimer atoms as a function of bias; note that the behavior is not linear.}
\label{bondlength}
\end{figure}

In the next set of simulations we relaxed the forces on a copper dimer in between the the plates of the capacitor, and determined the bond length as a function of bias. We used a solution probe coupling strength $\alpha$ of $0.01$ for all of the simulations: this is applied to the dimer and the atoms in the plates near the surfaces facing the opposite plates. In Fig. \ref{bondlength:sub1} we observe a quadratic dependence of the bond length with respect to the applied bias.

We can understand the dependence of bond length on applied bias directly from the TB model. Our symmetric dimer has one s-orbital per site with zero onsite energy, contains $N$ electrons, a hopping matrix element $v(r) = -v_0 (r_0/r)^q$ and a pair potential $\phi(r) = \phi_0 (r_0/r)^p$. This gives a total energy $E$ which has the form:
\begin{equation}
E(N,r) = -Nv_0\left(\frac{r_0}{r}\right)^q + \phi_0\left(\frac{r_0}{r}\right)^p
\label{E_N}
\end{equation}
Here, $r_0$ is a reference distance, and $v_0$ and $\phi_0$ the values of the corresponding hopping matrix element and pair potential respectively. These values, plus the powers $p$ and $q$, are taken from Sutton {\it et al.} \cite{Sutton2001}.

Because the dimer shares its electron chemical potential with the two plates, its value, and hence $N$, depends on the bias. There will be net charge at non-zero bias, as can be seen from the following argument. For the empirical TB model used here, the atom is assigned a core charge of 0.48608. This leads to partial filling of the dimer bonding orbital, which in turn means the bonding orbital is pinned to the reference chemical potential (an analysis of the DOS can be found in section IV C). The pinning of the orbital requires the injection of slightly too many electrons into the dimer, producing a negative base charge in the dimer atoms.

The equilibrium bond length $r=z$ corresponds to zero derivative of the energy with respect to changes in the bond length. This leads to the following dependence of the bond length on bias:
\begin{equation}
\frac{\Delta z}{z_0} = \left(\frac{N(0)}{N(V)}\right)^{1/(p-q)}-1
\label{dz}
\end{equation}
where $z_0$ is the equilibrium bond length at zero bias, and $\Delta z = z(V) - z_0$. The results of this model are compared with the results from the simulation in Fig. \ref{bondlength:sub1}, and excellent agreement is seen. Note that the variation with bias is nearly quadratic, and seen by the fitted trend line.

We now determine the polarizability $\alpha_E$ of the dimer from the variation of the charge difference between the two atoms with bias. In  a linear approximation we have:
\begin{equation}
\Delta Q = \frac{2\alpha_E V}{z_0 d}
\label{d_q_pol}
\end{equation}
where $\Delta Q$ is defined as $Q_{\mathrm{left}} - Q_{\mathrm{right}}$, where $Q_{\mathrm{left}}$ is the charge on the left atom, and $Q_{\mathrm{right}}$ is the charge on the right atom. As the field direction is right to left, one would expect $\Delta Q$ to be positive, which is indeed in agreement with the behavior seen in Fig. \ref{bondlength:sub2}. However, the variation is clearly non-linear, which might be a consequence of the net charge of the dimer: this charge leads to an energy penalty when more electron density is added to one of the atoms.

As the dependence of $Q$ on $V$ is non-linear, we choose the initial gradient $\left.\frac{\mathrm{d}Q}{\mathrm{d}V}\right|_{V=0}$ of the quadratic fitted to the data to estimate the polarizability. The separation $d$ between the two plates is $14.388$ \r{A} and $z_0$ is $2.15005$ \r{A}, yielding an initial polarizability of $37.03$ a.u. (Note that 1 a.u. corresponds to $1.648777 \times 10^{-41}$ C m$^2$ $V^{-1}$). Experimental polarizability values only exist for neutral copper atoms, so we can only compare the order of magnitude of the determined $\alpha_E$ to atomic values \cite{Ma2015,amerigo1995,Neogrady1997,Roos2005,schwerdtfeger1994}. Calculated values range from 40.7 a.u. to 58.7 a.u., with $58.6 \pm 4.7$ a.u. being the experimental value \cite{Ma2015}. The smaller value for the dimer relative to the free atom could well be a result of the minimal basis set used, which prevents polarization of individual atoms. However, the order of magnitude is the same.

There is one important consequence of the net charge on the dimer we have not yet considered. After relaxation, the dimer is only in a metastable state. If relaxed for long enough, it would eventually be pulled towards the positively charged plate. This was indeed what we observed upon increasing the number of relaxation steps.

\subsection{Density of States}

\begin{figure}[h]
\includegraphics[width=\columnwidth]{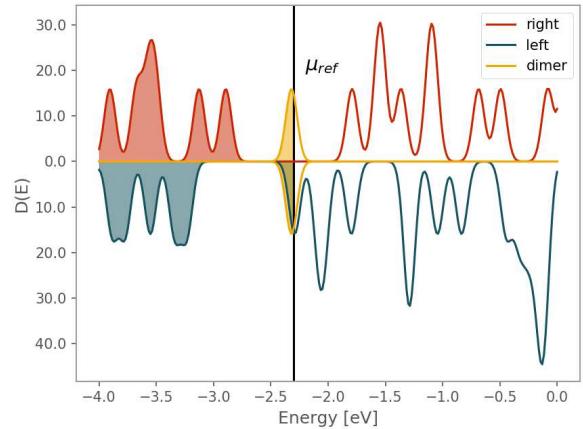}
\caption{The density of states for the setup shown in Fig. \ref{setup:sub2} after relaxation of the atomic forces. The upper (red) plot is for the right plate, while the lower (grey) plot is for the left plate. The dimer state is shown in yellow. The solution probe coupling strength is $\alpha = 0.01$, and the applied bias is 0.408 V. To make the plot smoother, we have applied a gaussian smearing of 0.05 eV.}
\label{dos}
\end{figure}

As previously mentioned, due to the empirical TB parametrization \cite{Sutton2001} used, the bonding orbital of the dimer is pinned to the reference chemical potential, which leads to high sensitivity of the number of electrons within the dimer to the bias, and results in a net negative base charge of the dimer in all of our simulations. The density of states plot in Fig. \ref{dos} shows that the dimer's bonding orbital is indeed pinned to the reference electrochemical potential, and is approximately half filled. One can also see the expected relative shift in density of states of the right and the left capacitor plate produced by the applied bias.

\subsection{Molecular Dynamics with Hairy Probes}

\begin{figure}[h]
\includegraphics[width=\columnwidth]{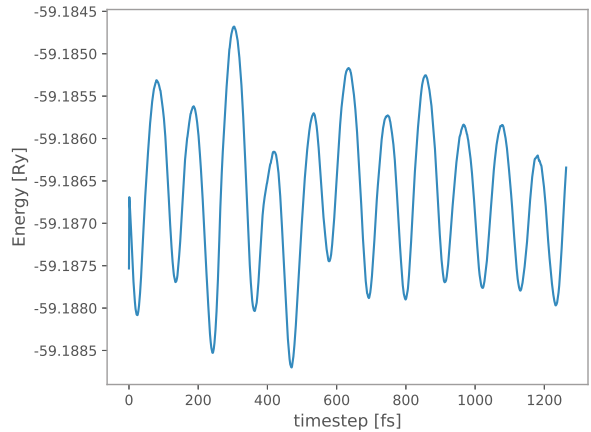}
\caption{The variation of energy trace with time from the molecular dynamics simulation of the system shown in Fig. \ref{setup:sub2}. The solution probe coupling strength is $\alpha = 0.01$, the bias is $0.408$ V, and the initial temperature of the atoms is 300 K.}
\label{trace}
\end{figure}

We conclude by considering a molecular dynamics simulation starting from the relaxed system, with an applied bias of 0.408 V. Our main concern is to  see how the energy depends on time, as this is a good indicator of how close our atomic forces are to derivatives of the anzatz for the energy with respect to atomic displacement. The  total energy evolution of the system over time is shown in Fig. \ref{trace}. We used a time step of 1 fs and an initial atomic temperature of 300 K. Note that the atoms to which the main probes have been attached (but not the solution probes) are not allowed to move. The fluctuations in energy shown in Fig. \ref{trace} are on the order of 0.001 Ry, with no systematic drift. This shows that the limiting case of Hairy Probes provides forces that are close enough to derivatives of the energy anzatz for us to perform MD simulations with open boundaries. This simple example simulation can be performed on a PC within a few hours, which opens the possibility of simulating larger electrochemical systems within a reasonable time frame.

\section{Conclusions}

Here we have presented a simple open boundary scheme for electrons that is appropriate for electrochemical simulations in which any electric current is the result of the diffusion and drift of charged ions in solution. It is derived as the limit of the Hairy Probes formalism in which the coupling of the probes to the system is weak. By means of a simple TB model of a capacitor between whose plates a dimer has been placed (the simplest approximation to an electrochemical cell we could envisage), we have shown that the forces are accurate enough, and the method efficient enough, that MD can be performed this way. This opens up a route to performing electronic structure simulations of non-equilibrium electrochemical processes at the atomic scale.

\begin{acknowledgments}
MGZ gratefully acknowledges funding from the EPSRC (EP/L015579/1) received through the Centre for Doctoral Training in the Theory and Simulation of Materials. APH acknowledges support from the Thomas Young Centre under grant TYC-10, and from the Leverhulme Trust under grant RPG-2014-125.

\end{acknowledgments}
\raggedright
\bibliography{paper.bib}
\end{document}